\documentclass[journal]{IEEEtran}
\usepackage{psfig,amssymb,amsmath,amsfonts,amsthm,multicol}

\begin{document}

\title{Wireless Network Coding via Modified 802.11 MAC/PHY: Design and Implementation on SDR}

\author{Mohammad~H.~Firooz,~\IEEEmembership{Student~Member,~IEEE,}
        Zhiyong~Chen,~\IEEEmembership{Student~Member,~IEEE}\\
        Sumit~Roy,~\IEEEmembership{Fellow,~IEEE,}
        and~Hui~Liu,~\IEEEmembership{Fellow,~IEEE}% <-this % stops a space
\thanks{Manuscript received December 08, 2011; revised March 02, 2012.}
\thanks{Mohammad H. Firooz and Sumit Roy are with the Department
of Electrical Engineering, University of Washington, Seattle,
WA, 98195 USA, e-mail: \{firooz,sroy\}@u.washington.edu.}% <-this % stops a space
\thanks{Zhiyong Chen and Hui Liu are with the Department of Electrical Engineering, Shanghai Jiao Tong University, Shanghai, 200240 China, e-mail: \{zhiyongchen,huiliu\}@sjtu.edu.cn.}% <-this % stops a space
\thanks{This work was supported in part by a grant from NSF under ECCS 0801997.}
}

\maketitle

\begin{abstract}
Network coding (NC), in principle, is a Layer-3 innovation that improves network throughput in wired networks for multicast/broadcast scenarios. Due to the fundamental differences between wired and wireless networks, extending NC to wireless networks generates several new and significant practical challenges. Two-way information exchange (both symmetric and asymmetric) between a pair of 802.11 sources/sinks using an intermediate relay node is a canonical scenario for evaluating the effectiveness of Wireless Network Coding (WNC) in a practical setting. Our primary objective in this work is to suggest \emph{pragmatic and novel} modifications at the MAC and PHY layers of the 802.11 protocol stack on a Software Radio (SORA) platform to support WNC and obtain achievable throughput estimates via lab-scale experiments. Our results show that network coding (at the MAC or PHY layer) increases system throughput-typically by $20-30\%$.
\end{abstract}
\section{Introduction}
The exponential growth of multimedia applications has resulted in current 3G cellular networks reaching (and exceeding, in the near future) available network capacities. As a result, communication engineers must find newer ways to continue to increase aggregate throughput while preserving Quality of Service (QoS). \emph{Cross-layer} approaches that seek to optimize aggregate network throughput based on adapting parameters from the physical or MAC layers have proven to be effective in this regard. Recently, Network Coding (NC) has attracted researchers' attention as another promising innovation in this context.

\subsection{Network Coding in Wired Networks}
Network coding was initially proposed as a distributed mechanism for achieving the multicast theoretic (max-flow, min-cut) capacity in wired networks. In wired multicasting, information is sent from a set of source nodes to a set of destination nodes over a multihop network where the intermediate nodes merely forward their received packets via a pre-determined look-up table (routing). Ahlswede et al., in \cite{ahlswede2000network} suggested the innovative notion of {\em coding} on layer-3 packets instead of look-up forwarding on specific outgoing links, and showed that network throughput can be increased.

In a network employing NC, routers perform a (random) linear combination of incoming layer-3 packets and broadcast the result to all its neighbors. Randomized linear network coding schemes were shown to be sufficient in achieving the information theoretic max-flow, min-cut bound on network capacity \cite{li2003linear}. Necessary and sufficient conditions for the design of such random linear codes were provided by Koetter et al. \cite{koetter2003algebraic}. While the concept of NC was developed for the network (IP) layer, it has often been implemented in practice at higher layers, such as the transport or application layers \cite{wang2007r2,firooz2010link,hamedICC}.

A fundamental reason as to why network coding is beneficial is based on the premise of {\em simultaneous transmission} from several (source) nodes to a single (receive) node. While this is feasible in a wired network whereby concurrent transmissions are deemed `orthogonal', a multi-hop {\em wireless} network is quite different. Wireless is a {\em shared} medium (at least for nodes within a common transmission range) and there is no natural spatial orthogonality.
Thus wireless multihop networks have relied on other forms of orthogonality -- in time (TDMA) or frequency (FDMA) -- to achieve interference-free transmission. Wireless Network Coding (WNC) uses non-orthogonal transmissions that, nevertheless, allow recovery of multiple packets to enhance aggregate network throughput.

\subsection{Network Coding in Wireless Networks}
The broadcast nature of wireless (coupled with network topology) determines the nature of interference. Simultaneous transmissions in a wireless network typically result in all of the packets being lost (i.e., collision). A wireless network therefore requires a scheduler (as part of the MAC functionality)  to minimize such interference. Hence any gains from network coding are strongly impacted by the underlying scheduler and will deviate from the gains seen in wired networks \cite{sagduyu2008cross}. Further, wireless links are typically half-duplex due to hardware constraints; i.e., a node can not simultaneously transmit and receive due to the lack of sufficient isolation between the two paths.

Another important consideration is the impact of the wireless channel on a transmitted signal -- inclusive of channel attenuation which is assumed negligible on a wire, but may not be ignored in wireless. The received signal $y$ over a wireless link can be modeled in general as
\begin{equation}
y=h\,x+z,
\end{equation}
where $x$ is the transmitted symbol, $z$ is the additive noise sample at the receiver, and $h$ is the (narrowband) channel loss between the source and the destination. Some previous work on WNC has incorporated aspects of the features mentioned above. Omnidirectional source transmissions were modeled in \cite{lun2006minimum}, \cite{lun2008coding} as hyper-links with additional constraints that prevent nodes from transmitting and receiving packets simultaneously. Interference effects were incorporated by \cite{lun2008coding} for joint optimization of MAC and network flows, where successful transmission between a node pair is based on a signal-to-interference-plus-noise-ratio (SINR) threshold, thereby potentially allowing simultaneous successful reception at different receive nodes.

One of the potential applications of WNC is in multicasting. A decentralized formulation to
throughput optimization for the multicasting problem was introduced in \cite{lun2005achieving}\cite{lun2004network}. However, if additional objectives such as maximizing throughput subject to delay constraints are considered, then network codes must be jointly designed with MAC as in \cite{sagduyu2007joint,wu2005minimum}. Authors in \cite{asterjadhi2010toward} qualify the impact of random access MAC schemes (such as CSMA/CA) on performance of NC in an all-to-all data dissemination system.

Information exchange via wireless relays is another natural scenario for potential application of network coding on top of the MAC layer. In \cite{wu2004information}, the authors assume a deterministic MAC protocol and generalizes the canonical three-node scenario to the case with an arbitrary number of relays between two source nodes. The main issue with MAC layer NC is that any existing asymmetry in the system may cause performance degradation. In fact, as we will show in our experimental result in Section \ref{S:Exp_result}, even small asymmetry in source nodes power decreases system throughput. Hence, MAC layer NC is naturally suited for symmetric transmission rates between any node pair. Since NC at MAC layer directly operates pointwise on the information symbols in the two packets from the respective sources, equal size packets (hence equal rates) are necessary.

The efforts to extend the idea of MAC layer NC to asymmetric traffic include \cite{NC}\cite{FL}. For example, \cite{FL} proposes to interpret NC as a mapping of modulation constellation to match symmetric traffic. However, it has been shown that, even in these schemes, performance is significantly reduced due to a lack of symmetry \cite{JM, SC}. To deal with asymmetric traffic, implementing the XOR operation of NC at the antenna has been proposed; this has been dubbed as \emph{Analog Network Coding} (ANC) or \emph{Physical Layer Network Coding} \cite{katti2007embracing}. Recently, Chen et al. \cite{JM} proposed a new network coding scheme called Decode-and-Forward
with Joint Modulation (DF-JM). They show that DF-JM has the potential to
achieve the capacity for asymmetric or symmetric traffic. Simulation-based evidence of the gains from WNC for a cellular network, in terms of achievable rate regions for MAC-layer and PHY-layer network coding, was presented in \cite{xue2007mac,liu2008network}, based on the assumption of perfectly known channels at every user.

The above summary of wireless network coding literature suggests that little effort has been put forth into actually implementing WNC concepts in any lab-scale prototype and measuring performance in a real setting.  One might say that, consequently, WNC is a largely unproven idea in practice; our work is thus distinguished by demonstrating some of the first evidence of the effectiveness of WNC in a canonical relay network.

\subsection{Wireless Relay Networks: A Review}
\begin{figure}
\centering
\begin{tabular}{c}
\psfig{figure=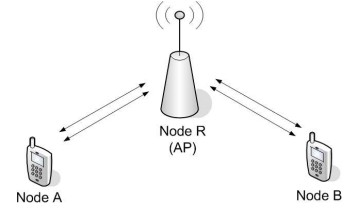,width=1.7in}
\end{tabular}
\caption{A relay network, Nodes $A$ and $B$ want to exchange data
through node $R$}
\label{F:relaynet}
\end{figure}

\begin{figure}
\centering
\begin{tabular}{c}
\psfig{figure=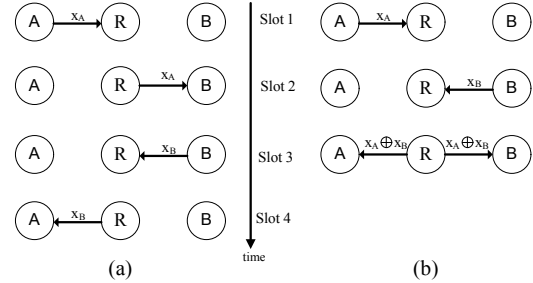,width=2.7in}
\end{tabular}
\caption{(a) 4-step and (b) 3-step information exchange using NC. In both scenarios, MAC protocol is TDMA.}
\label{F:3-4-step-relay}
\end{figure}

In this paper we investigate the challenges of implementing WNC in a simple relay network using a currently available software-defined platform, and suggest potential solutions. We present measured system throughput and compare it with suitable analytical results for benchmarking.
Consider a two-way (bi-directional) relay channel with two source/destination nodes (nodes $A$ and $B$), and one intermediate relay node (node $R$), as depicted in Figure \ref{F:relaynet}. By assumption, the relay node only forwards data and is not a source or destination. Nodes $A$ and $B$ are not within
transmission range of each other, and they require the relay node to communicate.
For a  half-duplex system, a baseline strategy for $A$ and $B$ is to exchange data via TDMA scheduling through node $R$. In the first time slot, node $A$ transmits data to $R$. In the next time slot, $R$
relays the received data to $B$. In the third time slot, node $B$ transmits data to $R$. Finally, in the fourth time slot, $R$ sends $B$'s information to $A$.
As depicted in Figure \ref{F:3-4-step-relay}-(a), two bits of
information would thus be exchanged in four time slots.

As already shown in \cite{larsson2006coded} and summarized in Figure~\ref{F:3-4-step-relay}-(b), it is possible to exchange two bits of information in three time slots by applying network coding in the MAC layer at the relay node, thereby achieving a 33\% throughput improvement relative to pure TDMA. In
the first time slot, $A$ sends its data $x_{A}$ to $R$. In the second
time slot, $B$ transmits its information, $x_{B}$ to $R$. In the third time slot,
node $R$ broadcasts $x_{A}\oplus x_{B}$. Since both nodes know their own
data, if they receive $x_{A}\oplus x_{B}$ they can extract their desired
information.

The above throughput enhancement requires an ideal TDMA MAC, which deterministically assigns a channel to each node in the network. However, one of most common wireless access networks (802.11) uses a random access MAC protocol based on carrier sensing (CSMA/CA) for time-sharing of the common channel. Besides the obvious advantages of a distributed protocol, such random access schemes are efficient and fair at low average traffic loads as in several data applications. Hence in this work, we first explore the challenges of implementing WNC over a system with random MAC protocol. The measured MAC layer throughput with NC shows significant gains in system throughput, close to the analytical predictions.

As discussed in the previous subsection, MAC layer network coding is based on {\em symmetric} traffic, whereas many data exchange scenarios are asymmetric by nature. Since the relay
node can be placed at varying distances between both sources, the data
rates transmitted by the relay node to both destinations are typically different.
As mentioned, PHY layer NC has been proposed to deal with asymmetric traffic. Unfortunately, there is relatively little attention paid in the literature to system implementation and actual performance measurement of applying NC at PHY layer. This is the second contribution of this paper.
We extend the concept of Decode-and-Forward with Joint Modulation (DF-JM) developed in \cite{JM}
(state-of-art PHY layer NC) to the asymmetric relaying problem considered here. In other words, we implement DF-JM to an OFDM-based (similar to 802.11a) PHY layer on a suitable test-bed
platform (described later) and measure actual  PHY layer NC throughput in a real-world application--a first, to the best of the authors' knowledge.

The current state-of-art of PHY NC
treats the addition of two (synchronized) analog signals at the relay antenna as the XOR operation over the information symbols. However, a closer model to reality for such a superposition should include the respective channel gains, i.e.,
\begin{equation}
y=h_1 x_1 + h_2 x_2 + z,
\end{equation}
where $x_i$ is the symbol from source $i$, and $h_i$ is narrowband channel gain between source $i$ and the relay. Note that, besides the requirement of synchronism, successful analog network coding also requires full knowledge of channel gains at the relay node.

\subsection{Contribution and Organization}
In summary, the main contribution of this paper is three-fold: a) to {\em design and modify} the CSMA-based MAC layer to support network coding, (b) to {\em develop} new ideas for PHY-NC for asymmetric traffic scenarios and {\em apply} the new ideas to OFMD-based PHY layer, and
(c) to {\em demonstrate} the utility of WNC via laboratory-scale experimentation using {\em commodity} wireless radios, notably 802.11a/g. Since much of NC theory has traditionally
presumed orthogonal time-scheduled MAC layer (such as TDMA), we believe that our results are some of the first to estimate the benefits from WNC for a CSMA/CA MAC.

The experiments were conducted using the Microsoft's new Software Radio (SORA) platform \cite{tan2011sora} and the open source 802.11a code.
SORA is a fully programmable software radio platform based on general purpose multi-core processors in commodity PC architecture, developed by Microsoft Research Asia (MSRA). With SORA, developers can implement and experiment with high-speed wideband wireless protocols (like IEEE 802.11a/b) using commodity general-purpose PCs \cite{zhang2010experimenting}.

The rest of the paper is organized as follows: In Section \ref{S:system_desc}, we discuss the challenges and possible solution of applying WNC both at MAC layer and at PHY layer of a two-way wireless relay network. Then in Section \ref{S:Imp}, we explain how to implement the proposed system in previous section in SORA platform running 802.11a. Experimental results about system performance and throughput are available at Section \ref{S:Exp_result}. Finally, the
paper concludes with reflections on future work in Section \ref{S:Conc}.

\section{System Description}\label{S:system_desc}
We propose to apply wireless network coding -- both at MAC and
at PHY layer - to a bi-directional relay network, as shown in Figure \ref{F:relaynet}.
 Our hardware platform,
SORA, implements 802.11a MAC and PHY layer. First, we explain how to
apply network coding on top of 802.11 MAC with minimum modifications. Then, we describe how the DF-JM scheme is implemented to support the new PHY layer NC concepts.

\subsection{802.11 MAC Layer Wireless NC}\label{SS:MAClayer}
The achievement of a $33\%$ improvement in idealized network throughput (displayed in Figure \ref{F:3-4-step-relay})
for the canonical two-node scenario communicating via a relay, is attained based on some key assumptions, notably: a) a scheduled MAC, such as
TDMA, with implicit node synchronization, and b) symmetric, constant-rate traffic, whereby source nodes have data to send in every slot. In the next two subsections,
we describe an implementation for the SORA with 802.11, which bypasses both of the above constraints.
To the best of our knowledge, there has been no demonstration of the benefits of NC in an 802.11
infrastructure network, where the Access Points act as natural relay between sources and sinks.
The known short-term unfairness due to carrier sense multiple access type MAC protocols such as 802.11 provides new challenges in implementing NC, and will be explored further \cite{garetto2008modeling}.
\subsubsection{Relay Node}
\begin{figure}
\centering
\begin{tabular}{c}
\psfig{figure=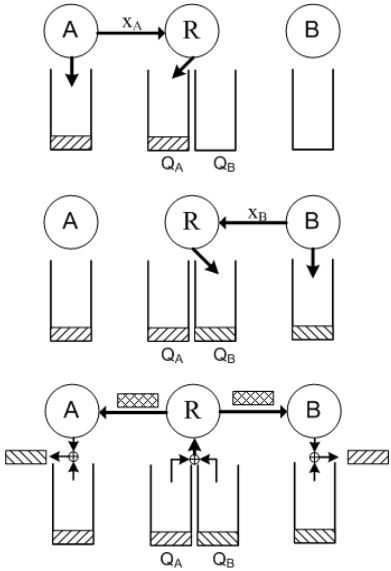,width=2.0in}
\end{tabular}
\caption{Proposed system architecture for MAC layer NC}
\label{F:proposed_system_buffer}
\end{figure}

\begin{figure}
\centering
\begin{tabular}{c}
\psfig{figure=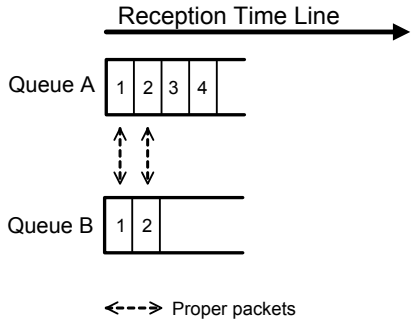,width=2.1in}
\end{tabular}
\caption{Simple illustration of proper packets}
\label{F:properPacket}
\end{figure}

The relay node $R$ (access point in 802.11) receives packets from both
sources, implements NC operation, and broadcasts the result.
Due to CSMA/CA channel access, the relay
may receive many packets from one source before it gets any packets from the other.
This necessitates two queues at the relay -- one each for packets from sources $A$ and $B$, respectively. Whenever
the relay receives a packet from one of
the sources, it looks at the other queue. If it finds the other queue
empty, it queues the packet and waits for packets from the other source to
arrive. The relay keeps queuing packets
from a source node until it receives one from the other source. Then it XORs one of the queued
packets from the first source and the (unique) \emph{proper packet} from the other, and broadcasts the result (Figure \ref{F:proposed_system_buffer}).
Roughly speaking, a packet from source $A$ ($B$) is called a proper packet of the packet from source $B$ ($A$), if they have the same reception order (Figure \ref{F:properPacket}).
Later, in Section \ref{SS:NCMAC_Imp}, we discuss the choice of such a proper packet in more detail.

The queue size at the relay node should be sufficiently large enough to achieve an acceptable packet drop rate. Further,
when the relay node receives a packet, it has to search the queue to find a \emph{proper packet} to be combined with the received one.  Accordingly, since searching a large queue is time and energy consuming, the queue size should also be sufficiently small enough to meet any NC delay constraint.
\subsubsection{Source/Sink Nodes}
Upon receiving a packet from the upper layer, the source node starts listening
to the channel. Whenever it finds the channel empty, it captures the channel
and starts transmitting. A copy of transmitted
packet is saved in its buffer, since it is needed to decode information from other sources as a result of NC. Whenever it receives a packet from the relay node, it checks its destination
address. If the packet was a broadcast packet, it fetches a \emph{proper packet} from
its buffer, calculates the XOR between the two, and sends the result to the upper layer.
\subsection{PHY Layer}
Upon receiving the broadcast messages from the relay node,
each destination decodes its intended message with its own signal as side
information. As such, the two-way relay can in general be regarded as
three separate slots, i.e., two sources send to relay node in slots 1 and
2 resp. and the relay broadcasts with side information in slot 3.
\subsubsection{Slot 1 and Slot 2}
In slot 1 ( slot 2), the information bits $w_{A}$($w_{B}$) are encoded and bit-interleaved
and applied to modulator,
generating the symbols $x_{A}$ ($x_{B})$ from the respective constellations
$\mathcal{M}_{A}(\mathcal{M}_{B})$, that is transmitted to the relay node.
Since the transmit rate mainly depends on the constellation size, we
consider BPSK modulation at both
sources for the $\emph{symmetric}$ relaying. For \textit{asymmetric}
relaying, $B$ transmits QPSK signals while $A$\ uses BPSK modulation or vice versa. During
the first time slot, the signal received at the relay node is thus given by
\begin{equation}
y_{AR}=\sqrt{P_{A}}h_{A}x_{A}+z_{AR},
\end{equation}
where $Y_{AR}$ and $Z_{AR}$ are the
received signal and the zero mean complex Gaussian noise of variances $%
\sigma _{AR}^{2}$ of the relay node, respectively. Likewise, the received
signal at the relay node during the second time slot is
\begin{equation}
y_{BR}=\sqrt{P_{B}}h_{B}x_{B}+z_{BR},
\end{equation}
where $h_{i}$, $i=A,B$, is the channel coefficient of
the link between the source/destination $i$ and the relay node, and
reciprocal channel can be assumed.
\subsubsection{Slot 3}
\begin{figure*}
\centering
\begin{tabular}{c}
\psfig{figure=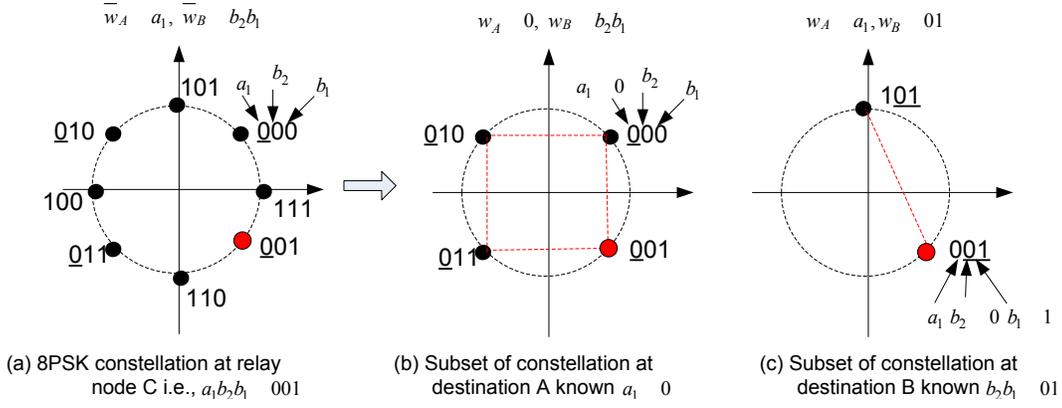,width=5.5in}
\end{tabular}
\caption{Demapping at both destinations for DF-JM schemes based on
optimal 8-PSK labeling }
\label{constellation}
\end{figure*}
During the third time slot, we apply the DF-JM scheme in the relay node to
broadcast the data. The relay node concatenates the
decoded information $\overline{w}_{A}$ and $\overline{w}_{B}$, into a new
sequence $\overline{w}_{R}=[\overline{w}_{A}\;\overline{w}_{B}]$, and then
encodes and modulates the resulting sequence jointly, regardless of the
sizes of the original messages. As an example, we consider the asymmetric
traffic with BPSK and\ QPSK constellations. Thus, we have $\overline{w}%
_{A}=[a_{1}]$ and $\overline{w}_{B}=[b_{2},b_{1}]$, where $a_{1}$,$b_{2}$
and $b_{1}$ denote the binary symbols; then $\overline{w}%
_{R}=[a_{1},b_{2},b_{1}]$. The transmitted signal from $R$ is
 $x_{R}=\mathcal{M} _{R}(\overline{w}_{R})$ by using the constellation $%
\mathcal{M} _{R}$. The corresponding received signals at stations $A$ and $B$ are respectively
 $y_{A}=\sqrt{P_{R}}h_{A}x_{R}+z_{A}$ and $y_{B}=\sqrt{P_{R}}%
h_{B}x_{R}+z_{B}$, where $z_{A}$ and $z_{B}$ are zero mean, complex
Gaussian noise of variances $\sigma _{A}^{2}$ and $\sigma _{B}^{2}$ .

With the help of known redundant bits $w_{A}$, node $A$ can
decode the desired bits $w_{B}$ using subset partitioning, i.e.
\begin{equation}
\ \widehat{w}_{B}=\underset{W_{B}}{\arg \min }\left\vert y_{A}-\sqrt{P_{R}}%
h_{A}\mathcal{M} _{R}([w_{A}W_{B}])\right\vert ^{2}.
\end{equation}
Likewise, $B$ can perform detection from the subset of the constellation
points based on the known sequence $w_{B}$. This is explained by the
example shown in Figure \ref{constellation} using 8-PSK constellation at the
relay node. The
relay node $R$ needs to forward $\overline{w}_{A}=[a_{1}]$ and $\overline{w}%
_{B}=[b_{2},b_{1}]$ to both destinations, where we assume that $a_{1}=0$
and $[b_{2} \; b_{1}]=[0 \; 1]$ without loss of generality. By applying the DF-JM scheme, the relay node $R$
combines them together as $\overline{w}_{R}=001$ and transmits $X_{R}=$($%
001$)$_{8PSK}$ using the 8-PSK constellation. Thanks to the fact that the
destination $A$ knows $a_{1}=0$, it only needs to consider the possible $%
[b_{2}\;b_{1}]$ from all the 8-PSK points for which the first bit\ is $0$, as shown
in Figure \ref{constellation}. Similarly, the estimated $a_{1}$ for
destination $B$ can be chosen from the points ($001$)$_{8PSK}$ and ($101$)$%
_{8PSK}$. These results implicates that \emph{signal detection can be
performed from a subset of the constellation points instead of the entire
constellation}.
\begin{figure}
\centering
\begin{tabular}{c}
\psfig{figure=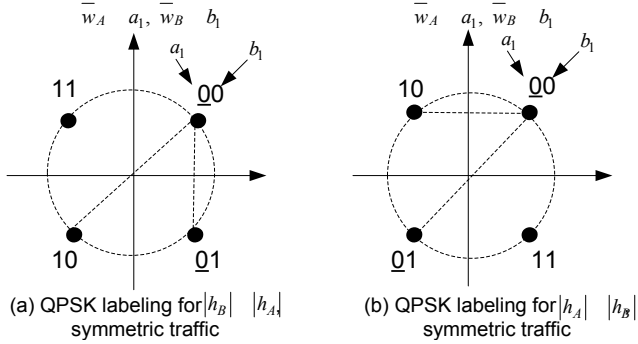,width=3.3in}
\end{tabular}
\caption{Optimal constellation labeling for QPSK constellation at the relay
node}
\label{QPSKconstellation}
\end{figure}
Therefore, we can see that for DF-JM scheme, the high level constellation is
used in the relay node, but low level constellation can be used for de-mapping
at sink nodes $A$ and $B$ by exploiting the side information. From
Figure \ref{constellation}, we also can see that if the constellation labeling
map is carefully chosen in the relay node, the intra-subset Euclidean
distance (i.e., ($001$)$_{8PSK}$ and ($101$)$_{8PSK}$) can be greatly
increased through the side information. It can be verified that such \
labeling for 8-PSK in Figure \ref{constellation} is optimal;
the optimal labeling for QPSK is shown in Figure \ref{QPSKconstellation}.
\section{SORA Implementation}\label{S:Imp}
The network stack of the legacy 802.11 implementation in SORA is depicted in Figure %
\ref{F:sora-structure} and consists of three parts \cite{tan2011sora}:
\begin{itemize}
\item PHY layer--this layer is OFDM-based and similar to 802.11a PHY layer.

\item MAC layer--this layer is simply a state machine modeling CSMA/CA, the core component of Distributed Communication Function (DCF) in 802.11a/b/g.

\item Link layer (LL)--this layer is responsible for interfacing with TCP/IP layer. When a node receives a MAC frame, this layer decides what to do with it. If the packet is addressed to this receiver, LL passes it to TCP/IP layer; otherwise, the packet is dropped.
\end{itemize}
Next, we describe how this stack was changed to support MAC and PHY network coding as described above.
\subsection{NC at MAC layer}\label{SS:NCMAC_Imp}
\begin{figure}
\centering
\begin{tabular}{c}
\psfig{figure=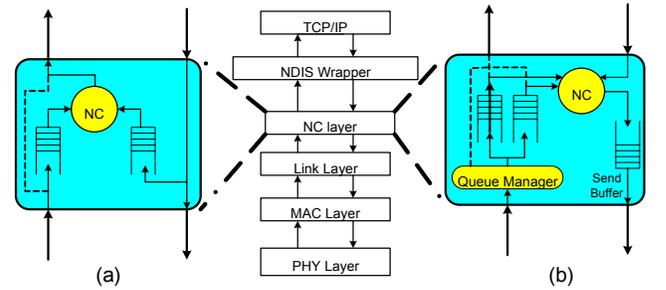,width=3.3in}
\end{tabular}
\caption{Proposed 802.11 network stack for (a) source/sink (b) relay node(s) in SORA}
\label{F:stackPort_source_relay}
\end{figure}

To apply network coding at the MAC layer, the legacy PHY
and MAC layers implemented in SORA architecture remain untouched. All changes are applied to Link layer via addition of an NC layer, dedicated to NC operation. As mentioned in Section \ref{SS:MAClayer}, this layer
is different for source/sink nodes and the relay node. However, they both need a
network coding unit, which simply XORs data on its input ports. Figures \ref{F:stackPort_source_relay}-(a) and \ref{F:stackPort_source_relay}-(b) show the new stack
protocol for relay and source/sink nodes, respectively.

\subsubsection{Packet format}
When a source receives a packet from the relay node, it knows that the packet is a result of XOR combination of one of its own packets and the desired packet transmitted from the other source. But it may already have sent a number of packets before receiving any from the relay. Hence it needs to know which one of its packets has been used for encoding at the relay node. To solve this problem, Chou et al. proposed the concept of `generation' for a packet \cite{chou2003practical}. Each packet contains a metadata field in its header representing its \emph{generation}. Only packets from the same generation can be combined. Each source has a \texttt{counter} which shows its current generation number. At the beginning, both sources reset their generation counter to zero. Whenever they transmit a packet, they insert the current generation number in the related field in a packet header and increment the \texttt{counter} by one.

The relay node only combines two packets from two sources if they are from the same generation. If it receives a packet from source $A$ ($B$) with generation number $n$, it first searches in the buffer dedicated to source $B$ ($A$) looking for a packet with the same generation number $n$. At that point, one of the following two scenarios will ensue:
\begin{itemize}
\item There is a packet with generation $n$. In this case the relay node will combine the two packets and broadcast the result. Then, it look at the buffer dedicated to $A$ ($B$) and deletes any packet from generation $n$.

\item The buffer is empty or there is no packet from generation $n$ - in this case, relay node will save the packet in the buffer dedicated to $A$ ($B$). If there is already a packet from generation $n$ in $A$ ($B$) buffer, it will be rewritten by the new one.
\end{itemize}
When a source receives a packet, it looks at its generation number. It fetches a packet with the same generation number from its buffer, fulfills network coding operation on the two packets, and sends the result to the upper layer.

Note that the variable \texttt{counter} that keeps track of a packet's generation is simply a register in a finite field, which means that the value contained is bounded and will reset after a while. Therefore, when two packets from generation $n$ are combined, there is a possibility that they actually belong to different generations. The probability of this occurring depends on the size of the \texttt{counter} and is minimized by a sufficiently large value of generation ID (or \texttt{counter}). On the other hand, adding generation ID to the packet header increases packet overhead, representing a tradeoff in choice of generation ID field size.
\begin{figure}
\centering
\begin{tabular}{c}
\psfig{figure=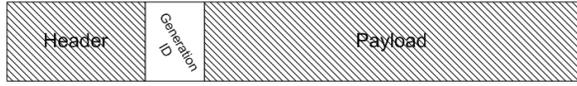,width=3.0in}
\end{tabular}
\caption{Field for generation number in packet header}
\label{F:generation_packet}
\end{figure}
\subsubsection{Source/Sink Nodes}
Each source/sink node has two major threads, one for receiving a packet (sink part) and  one for transmitting  (source part). To implement the MAC layer NC, the receiving thread must know about the packet flow in the sending thread, i.e. the receiving thread needs to know exactly which packets have been transmitted to complete decoding. For that reason, as discussed in Section \ref{SS:MAClayer}, there is a shared \texttt{buffer} with bounded length \texttt{BUFFER\_SIZE} between the receiving thread and sending thread in each source node.

When a packet is transmitted, the sending thread puts a copy of the packet in the \texttt{buffer}. If the buffer is full, it will overwrite the oldest packet in the buffer. If the source receives a packet, the receiving thread looks at its generation ID and fetches a packet, if any, from the \texttt{buffer} with the same generation number from its buffer. Then the node XORs the two packets (received packet and one from the buffer) and sends the result to the upper layer. If there is no packet with that generation ID, the receiving thread sends the packet to the upper layer. If the packet has undergone NC (XOR at the relay node), then the CRC check at the upper layer fails and the packet is dropped; only a non-NC directly from the other source is accepted. Note that since the \texttt{buffer} is shared between two threads, lock protection between the threads is needed \cite{silberschatz1994operating}.

\subsubsection{Relay Node}
\begin{figure}
\centering
\begin{tabular}{c}
\psfig{figure=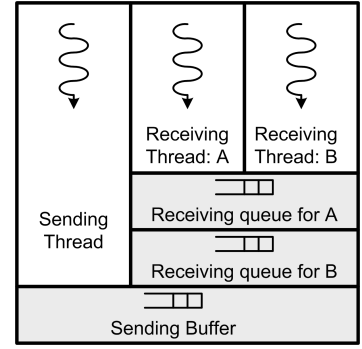,width=1.8in}
\end{tabular}
\caption{Multithreading and shared queue configuration at the relay node}
\label{F:threads}
\end{figure}

The relay node has three major threads: the receiving thread for $A$, the receiving thread for $B$, and the sending thread (Figure \ref{F:threads}). Two receiving threads share two buffers as depicted in Figure \ref{F:proposed_system_buffer}-- one for each source. When the relay receives a packet from $A$ ($B$), its dedicated thread wakes up and extracts the generation number from the received packet. It then searches inside the buffer dedicated to $B$ ($A$) for a packet with the same generation ID. If there is such a packet, the relay node combines the two packets, signals the sending thread, and hands the resulting packet to it to broadcast. If there is no packet with the same generation ID as the received packet, it is saved in a buffer dedicated to $A$. If the buffer is full, the oldest packet would be fetched out, handed to the sending thread for broadcast, and replaced with the new packet.

As depicted in Figure \ref{F:threads}, the three threads at the relay node share the \texttt{sending buffer}. Whenever a packet is ready to be sent, it is placed in the \texttt{sending buffer}. If the buffer is full, the oldest packet would be overwritten. The sending thread always monitors the \texttt{sending buffer}; while the buffer is not empty, the sending thread fetches the oldest packet and sends it to the PHY layer.
\subsection{NC at PHY layer}

\subsubsection{Frame Structure}
\begin{figure}
\centering
\begin{tabular}{c}
\psfig{figure=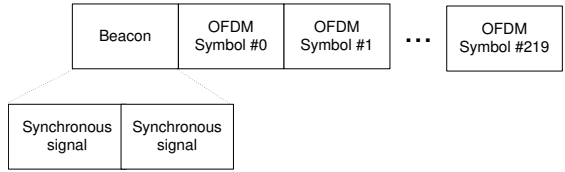,width=3in}
\end{tabular}
\caption{PHY symbol frame structure in OFDM system} \label{frame}
\end{figure}

\begin{figure}
\centering
\begin{tabular}{c}
\psfig{figure=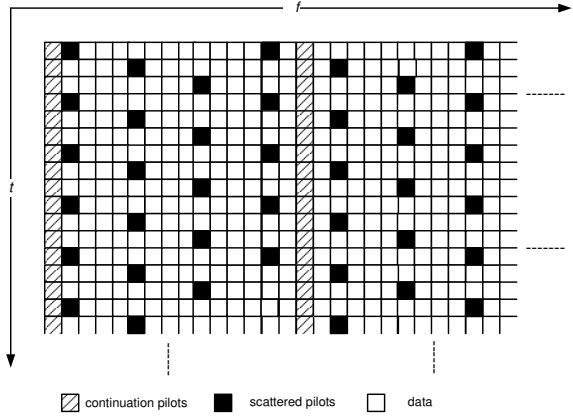,width=3.0in}
\end{tabular}
\caption{Signal types in time-frequency grid}
\label{signal_type}
\end{figure}

As shown
in Figure \ref{frame}\footnote{%
Note that a basic OFDM PHY is implemented here to support the
experiments, which can be extended to the PHY frame in 802.11a.}, the physical layer is based on
11.52 ms duration frames, comprising of beacon and data components. The beacon
consists of two identical band limited pseudo random signals, mainly
used for frame synchronization and frequency offset estimation. The
data signal is composed of 220 OFDM symbols. Each OFDM symbol
includes 256 subcarriers, but only 198 subcarriers is used, which
includes 168 data subcarriers, 12 continuation pilots and 18
scattered pilots. The three
types of signals are transmitted over the OFDM time-frequency grid as in Figure \ref{signal_type}.
The information for physical layer such as modulation and coding
scheme and the type of network coding are included in continuation
pilots. The continuation pilots can also serve for
frequency synchronization and phase tracking. The rectangularly
distributed scattered pilots
serve as reference for channel estimation, and the power of
continuation pilots and scattered pilots are set to be 3 dB higher
than data signal. Table \ref{T:wifiParam} summarizes the detailed PHY parameters
 that supports BPSK, QPSK, and
8-PSK modulation schemes for different relay scenarios.

\begin{table}
\caption{OFDM PHY parameters}\tabcolsep 5mm
\par
\begin{center}
\begin{tabular}{ll}
\hline Parameter & Value \\ \hline
Frequency band & 2.422 GHz \\
signal bandwidth & 4.254 MHz \\
subcarrier spacing & 21.484 KHz \\
Symbol duration (data) &  46.546 $\mu$s\\
Guard interval duration & 5.8182 $\mu$s \\
Frame length & 11.52 ms \\
LDPC length & 9216 \\
Code rate & 1/2 \\ \hline
\end{tabular}%
\end{center}
\label{T:wifiParam}
\end{table}
\subsubsection{Packet Relaying Processing}
\begin{figure}
\centering
\begin{tabular}{c}
\psfig{figure=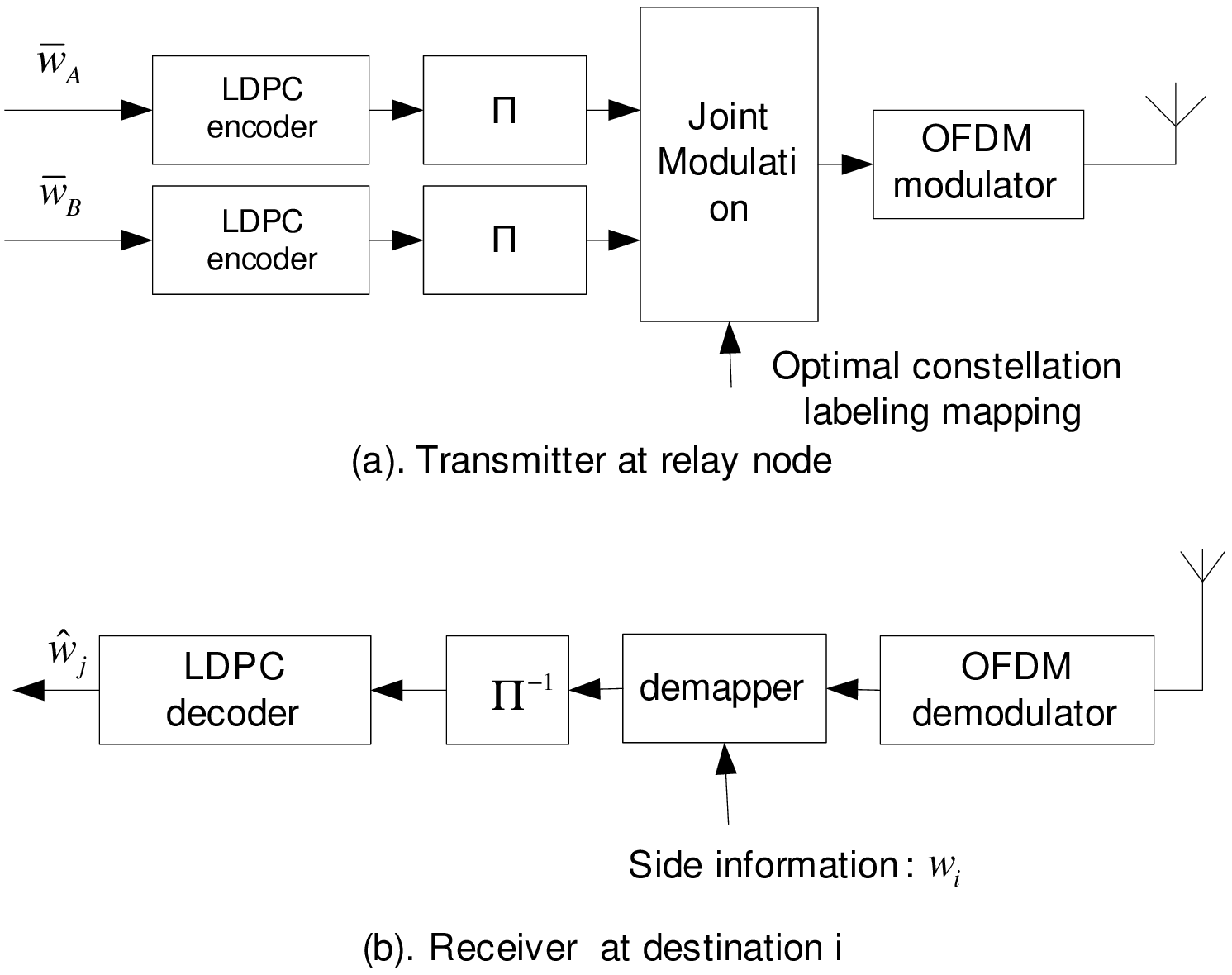,width=3.0in}
\end{tabular}
\caption{Signal processing in the relay node and destinations (source/sink nodes) for PHY layer NC}
\label{processing}
\end{figure}

Since the DF-JM is employed in the relay node, we extend the OFDM PHY
with DF-JM scheme; the associated signal processing for the transmitter
and receiver is shown in Figure \ref{processing}. At the third
time period when the relay node becomes a transmitter, two LDPC
encoders will be used to encode the information from the two
sources.
Based on channel condition, three types of modulation are considered in this paper: BPSK, QPSK and 8-PSK packets.
For symmetric relaying, nodes $A$ and $B$ transmit the BPSK packet to
the relay node $R$, and then QPSK packet can be transmitted by the
node $R$ using the DF-JM scheme. When the node $R$ is placed near by node $B$, the link between $B$ and $R$ can
support QPSK transmission, allowing the node $R$ to
transmit the 8-PSK packet based on DF-JM scheme in the third time
slot. For comparison, optimal labeling and Gray mapping for QPSK and
8-PSK packets are introduced at the node $R$ and the
labeling information carried by the continuation
pilots.
\section{Experimental Results}\label{S:Exp_result}
\subsection{Sora Platform}
Software defined radios (SDR) have been attracting increasing attention recently. In an SDR system, components that are typically implemented in hardware (e.g., mixers, filters, amplifiers, modulators/demodulators, detectors, etc.) are instead implemented by means of software on a suitable hardware platform \cite{dillinger2003software}. Changing a component implemented in software is easier and faster, leading to consequent flexibility in modifying a communication system built on a SDR.

\begin{figure}
\centering
\begin{tabular}{c}
\psfig{figure=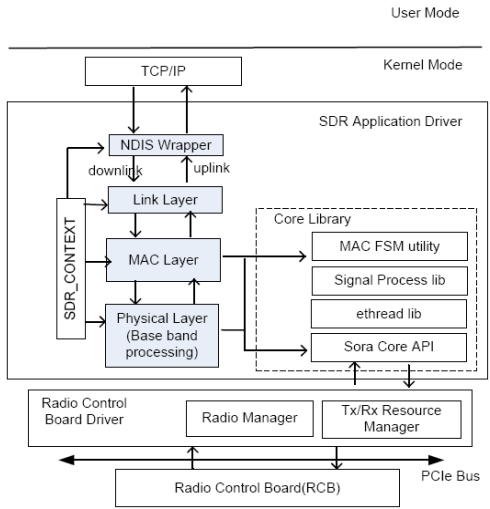,width=3in}
\end{tabular}
\caption{Hardware/Software structure in SORA \cite{tan2011sora}}
\label{F:sora-structure}
\end{figure}

Microsoft's Software Defined Radio  platform (SORA) consists of three fundamental components:  \cite{tan2011sora}:
\begin{itemize}
\item RF front-end
\item Radio Control Board (RCB)
\item SDR application driver
\end{itemize}
Figure \ref{F:sora-structure} illustrates the SORA architecture. The RF front-end represents the well-defined interface between the digital and analog domains. It contains analog-to-digital (A/D) and digital-to-analog (D/A) converters, and necessary circuitry for radio transmission. Since all signal processing is done by the software, the RF front-end design is rather generic. The RF front-end in SORA is interchangeable; in this work, we use the WARP radio board \cite{warp}.

The RCB is the new PC interface board for establishing a high-throughput, low-latency path for transferring high-fidelity digital signals between the RF front-end and PC memory. To achieve high system throughput, RCB uses a high-speed, low-latency PCIe bus, which supports a throughput of up to 16.7 Gbps with x8 model. The large on-board memory further allows caching pre-computed waveforms, adding additional flexibility for software radio processing.
Finally, the RCB provides a low-latency control path for software to control the RF front-end hardware and to ensure it is properly synchronized with the host CPU.

In SORA, SDR components are located in a separate upper-level driver, called an SDR Application Driver, where the customized Link layer, MAC and PHY are implemented. An SDR application driver accesses the hardware resource via the Sora Core API inside the Core Library. The core library implements the common functions of radio and other hardware resource management, like radio board configuration and control. Specially, it provides the necessary system services and programming support for implementing wireless PHY in a general-purpose
multicore processor. Several processing tasks with intensive computation complexity, such as the down-sampling,
Fast Fourier Transform (FFT) and LDPC decoder, are optimized by taking advantage of the data-parallelism with the
Single Instruction Multiple Data (SIMD) instruction sets of the CPU.

\subsection{NC at MAC layer: Experiments}
We use three i7 personal computers supporting PCIe. Each computer is equipped with a SORA board and a warp radio board as the RF front-end. To remove any hidden node problem, we place each node on a vertex of an equilateral triangle with a side length of 1 m \footnote{Estimating the impact of hidden node is out of scope of this paper.}. In this configuration, every node is in coverage range of the others. However, source nodes only accept packets from the relay node.

Each node is performing in 802.11a basic mode with a transmission rate of 1 Mbps. Each source contains a 135 MB file aiming to send it to the other one. Packets are 1 KB long; i.e., it takes roughly 8 ms to send a packet. Length of generation ID is 32-bit, which is a negligible overhead compared to the length of the packets. For the sake of simplicity, in this work, the same amount of memory is assigned to every queue in the system architecture discussed in Section \ref{SS:NCMAC_Imp}. In other words, all queues would have the same size in each experiment. Having different sizes for each queue and observing the role that plays on system performance is deferred to future work.

\begin{figure}
\centering
\begin{tabular}{c}
\psfig{figure=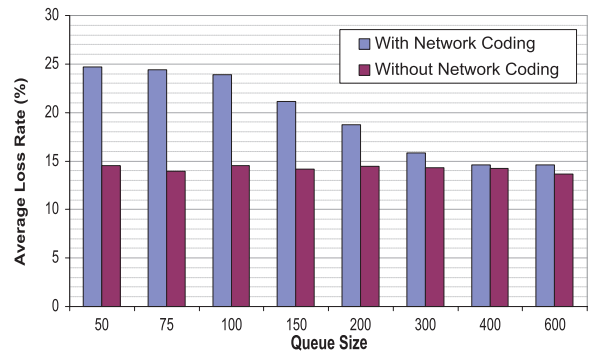,width=3.0in}
\end{tabular}
\caption{Average packet loss rate in two-way relay on 802.11 framework with and without NC at MAC layer}
\label{F:lossrate}
\end{figure}

\begin{figure}
\centering
\begin{tabular}{c}
\psfig{figure=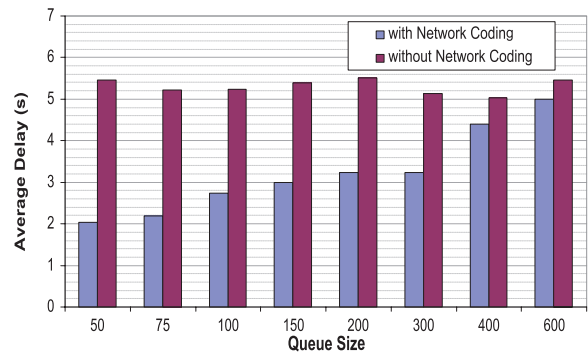,width=3.0in}
\end{tabular}
\caption{Average delay in two-way relay on 802.11 framework with and without NC at MAC layer. Delay axis has been scaled such that the packets are 1 second long (equivalently 1Mbit long).}
\label{F:aveDelay}
\end{figure}

Figure \ref{F:lossrate} compares packet loss rates in the relay network with and without network coding. As we discussed in Section \ref{SS:NCMAC_Imp}, using network coding increases the likelihood of losing packets. A solution to this problem would be increasing the size of the buffer. As one can see, having a large enough queue would overcome any packet loss caused by MAC layer NC. However, increasing queue size, as Figure \ref{F:aveDelay} shows, would increase the average delay in the system. In this instance, we define delay as amount of time needed for a packet to traverse from one source to the other. In this figure, we normalize the y-axis such that a packet length is 1 s. At the cost of increasing the incidence of packet loss and a little more computational complexity at the nodes, MAC layer NC decreases packet delay at low queue size. If one increases size of the queue to decrease system packet loss, that cancels out the advantage of shorter packet delay as in Figure \ref{F:aveDelay}.

\begin{figure}
\centering
\begin{tabular}{c}
\psfig{figure=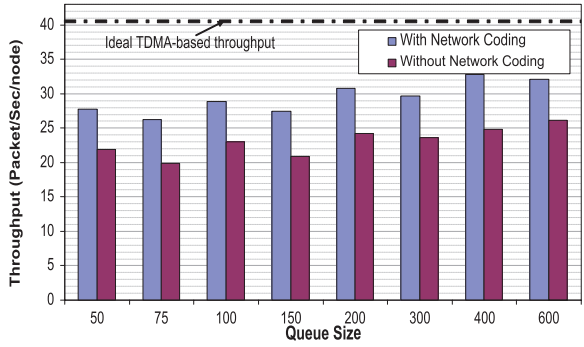,width=3.0in}
\end{tabular}
\caption{System throughput (per node) in information exchange for a 802.11 network when MAC layer NC is applied}
\label{F:throughput}
\end{figure}

System throughput (packet/s/node) is depicted in Figure \ref{F:throughput}, where throughput is the average number of successfully received packets at each node per second.
Comparing to baseline TDMA relay protocol (4-step without network coding), NC at MAC layer increases the throughput by about $20-30\%$. While system throughput for applying NC to CSMA/CA MAC is always less than $35$ packets/s/node, for ideal TDMA-based system, the throughput would be around $41$ packets/s/node. That means, when network coding is applied to MAC layer, randomness in capturing the channel causes nearly $15\%$ reduction in system throughput compared to an ideal deterministic TDMA MAC.

\begin{figure}
\centering
\begin{tabular}{c}
\psfig{figure=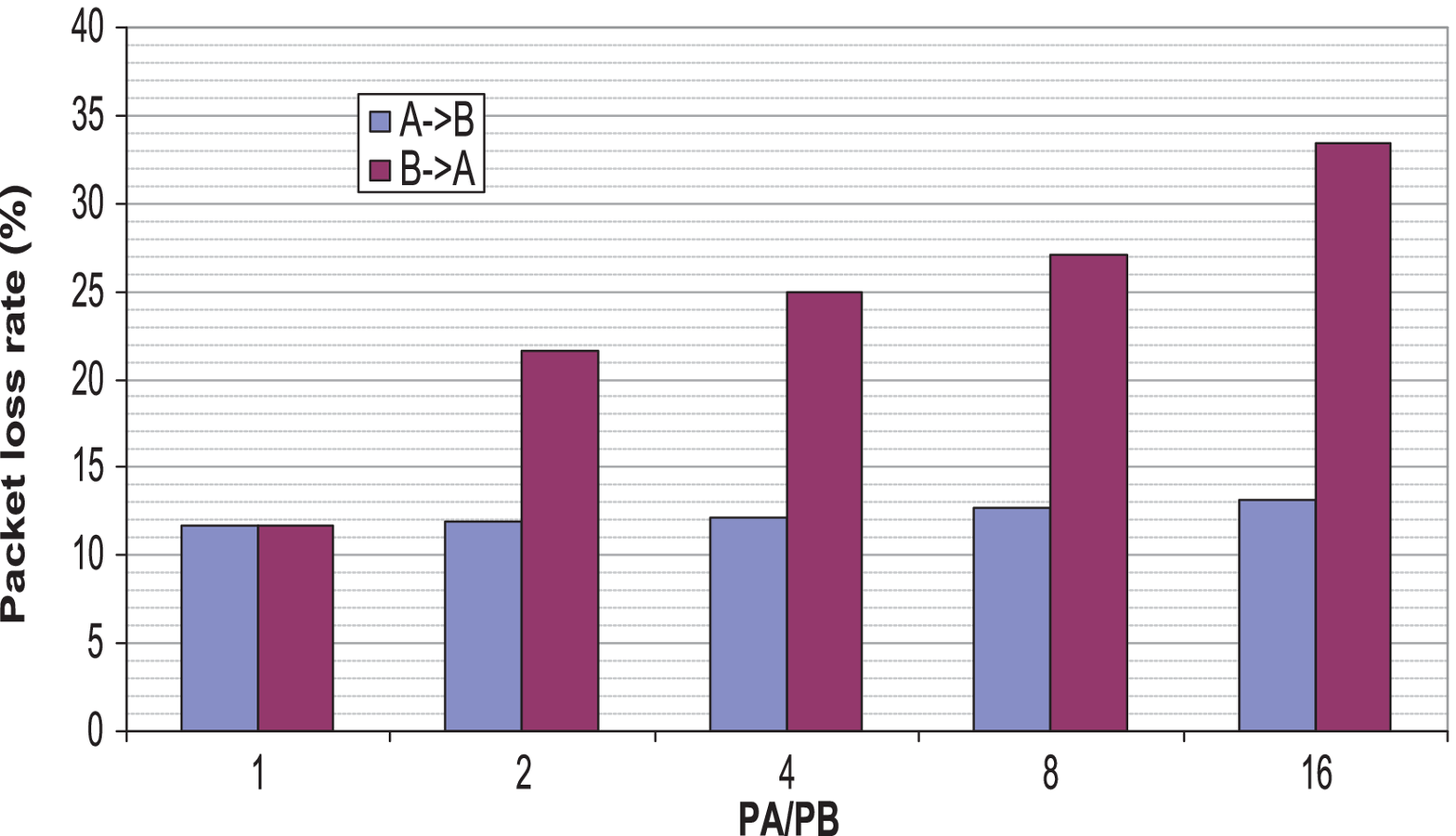,width=3.0in}
\\
(\textbf{a})
\\
\psfig{figure=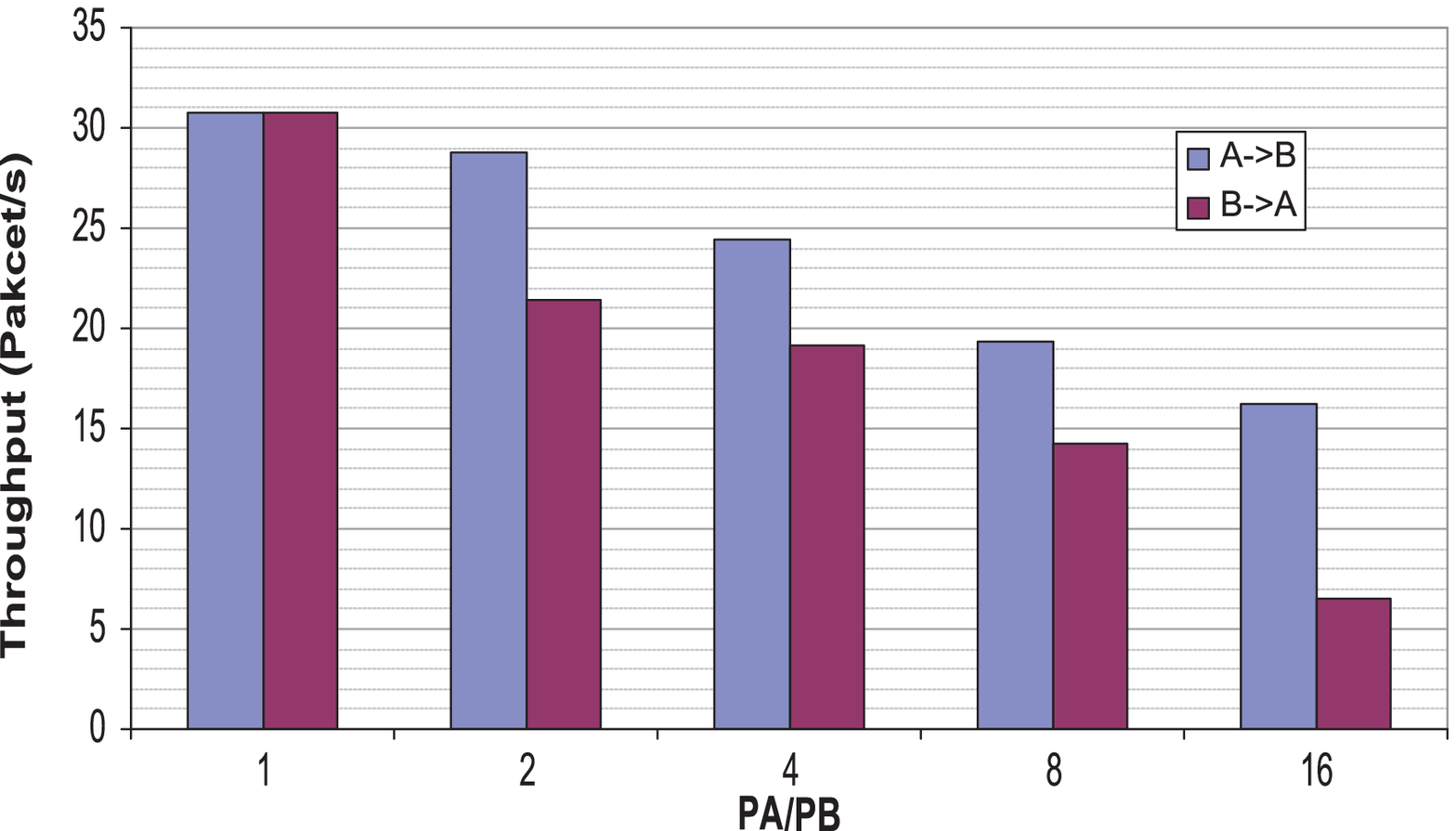,width=3.0in}
\\
(\textbf{b})
\end{tabular}
\caption{reducing transmission power at node $B$ to measure (a) packet loss rate (b) throughput of each node when MAC layer network coding is used}
\label{F:diffP}
\end{figure}

Finally, to test the sensitivity of MAC layer NC, we change one of the nodes' power and keep the other parameters of the system untouched. In each experiment, we halve the power of node $B$ and measure system throughput and packet loss rate for each source/sink in the system, as depicted in Figure \ref{F:diffP}. Figure \ref{F:diffP}-a presents the following results: when the power of node $B$ decreases, its packet loss rate increases while the packet loss rate of the other node (node $A$) remains almost the same. For MAC layer NC, the relay node needs to receive packets from both nodes in order to do the encoding (XOR the packets) and broadcast the result. Hence packets from node $A$ are buffered until a packet successfully arrives from node $B$. This means delay from $A$ to $B$ increases or, equivalently, throughput decreases. As one can see in Figure \ref{F:diffP}-b, having asymmetry in the system would decrease throughput of node $A$ as well. The reason for that can be justified as follows: buffer dedicated to node A at the relay node would be filled up because of the high packet loss rate from node $B$. As discussed in Section \ref{SS:NCMAC_Imp}, the relay node would start broadcasting uncoded packets from node $A$ to prevent buffer overflow. This means the sending buffer at the relay usually has some packets from $A$ to be sent. Therefore, when a packet is received from $B$ and gets encoded by the correspondent packet (i.e. same generation) from node $A$, the coded packet would spend some time in relay's sending buffer, increasing the delay from $B$ to $A$ and degrading throughput at $A$.
\subsection{NC at PHY layer}
As our experience in the previous subsection shows, unequal packet loss rate would decrease system performance. PHY layer NC is proposed to deal with any asymmetric traffic.
In this subsection, the performance of the demonstration platform with the novel PHY prototype is
shown. In the test set-up, the platform was implemented
on three PCs equipped with external antennas. By adjusting the transmit power
and the distance between transmit and receive antenna pairs, the channel
condition can be changed for different scenarios.

\begin{figure}
\centering
\begin{tabular}{c}
\psfig{figure=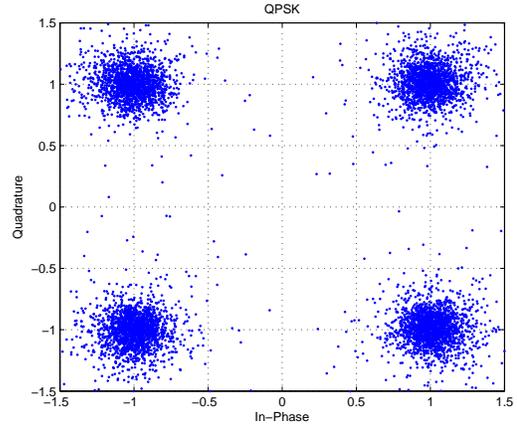,width=3in}
\end{tabular}
\caption{The received signal after channel
equalization in $A$ for symmetric relaying, SNR =16.0632
dB }
\label{QPSKC}
\end{figure}

\begin{figure}
\centering
\begin{tabular}{c}
\psfig{figure=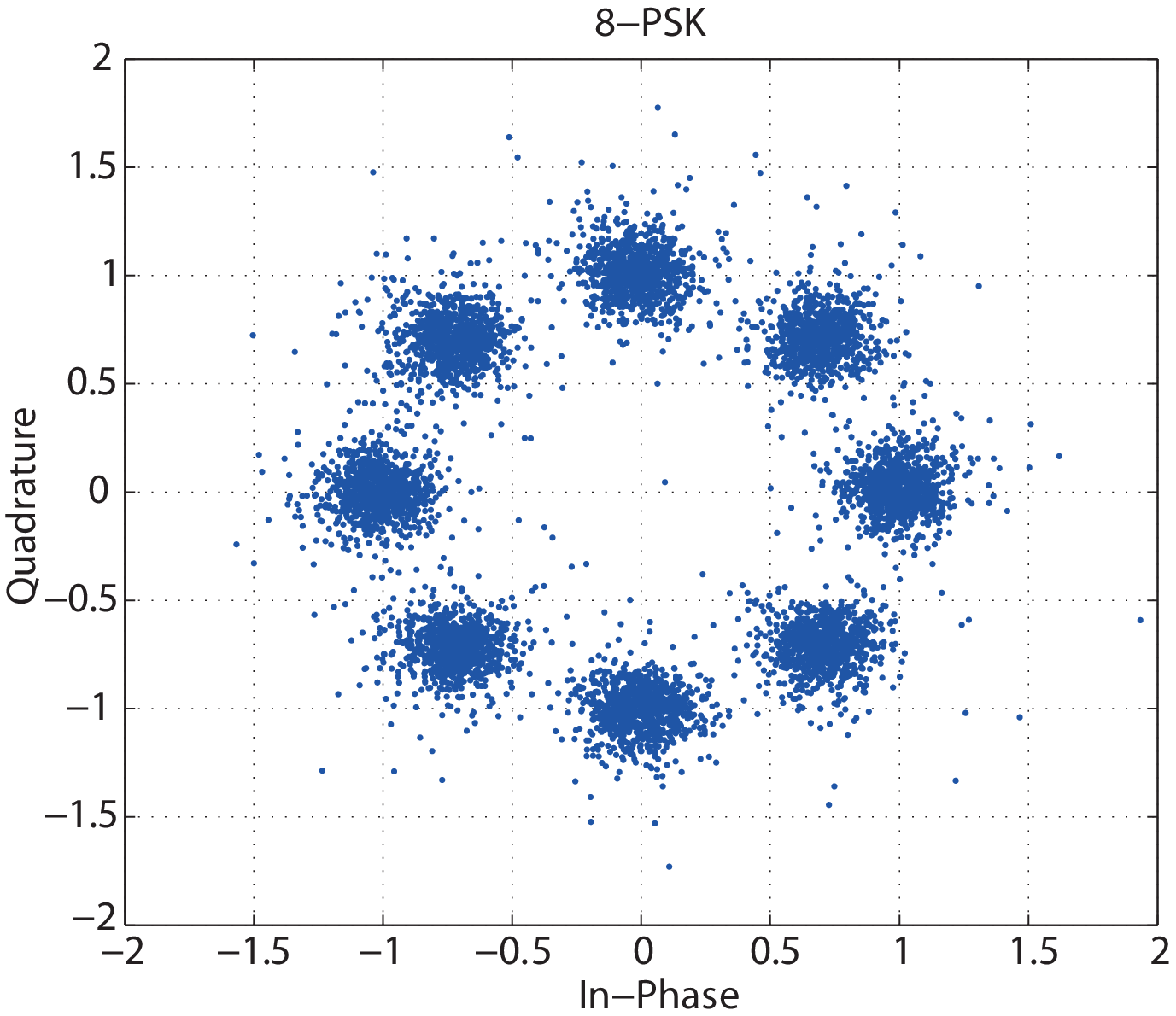,width=3in}
\end{tabular}
\caption{The received signal
after channel equalization in $A$ for asymmetric relaying, SNR =16.4238 dB}
\label{8PSKC}
\end{figure}

\begin{figure}
\centering
\begin{tabular}{c}
\psfig{figure=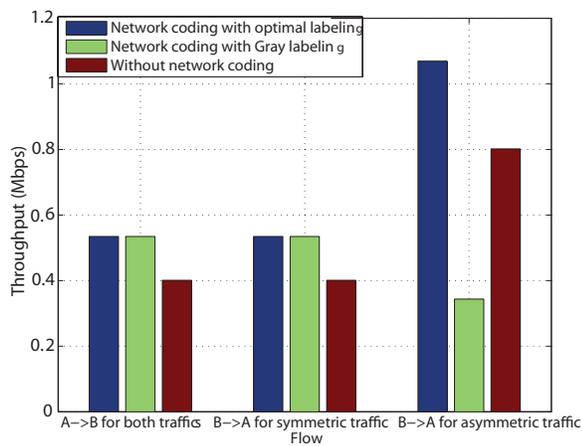,width=3in}
\end{tabular}
\caption{The end-to-end throughput (per flow) in information
exchange for an OFDM-based network when PHY layer NC is applied}
\label{throughput}
\end{figure}

In Figures \ref{QPSKC} and \ref{8PSKC}, we illustrate the
received signal after channel equalization at node A for the
two-way symmetric and asymmetric traffic. The end-to-end throughput
for both symmetric and asymmetric traffic corresponding to
different network coding schemes is presented in Figure
\ref{throughput}. It is observed that irrespective of symmetric
 or asymmetric traffic, a three-step information exchange scheme
based on DF-JM with optimal labeling significantly improves the network
throughput, compared to four-step information exchange scheme. It is
noteworthy that the optimal labeling provides a nearly $100\%$ gain over
the Gray mapping at SNR around 7dB for asymmetric relaying, but
does not have the same throughput improvement for symmetric relaying.
This behavior is reasonable because the SNR=7dB is\ enough to
support the transmission of QPSK signal without decoding error.

\section{Conclusion}\label{S:Conc}
We modified 802.11 MAC to support network coding on SORA platform from Microsoft, and conducted experiments to measure system performance (delay, packet loss rate and
throughput) in a three node with relay network. Our results showed MAC layer NC improves system throughput by twenty to thirty percent but with the cost of increasing packet loss rate.

Moreover, we explored one of the fundamental problems with MAC layer NC - its known sensitivity to symmetric traffic. System throughput dramatically drops when there the system supporting MAC layer NC is not symmetric.  We explained how PHY layer NC can be used to overcome this problem. We designed and implemented an OFDM-based PHY (similar to 802.11a PHY) to support NC. Our lab-scale experiment on SORA showed that for asymmetric traffic, our implemented PHY layer NC increases throughput by thirty percent comparing to traditional 4-step TDMA exchange.
%\section*{Acknowledgment} This work was supported in part by a grant from NSF under ECCS 0801997.
\bibliographystyle{IEEEtran}
\bibliography{IEEEabrv,overall}

\end{document}